\documentclass[aps,pra,twocolumn,superscriptaddress]{revtex4}
\usepackage{amssymb}
\usepackage{amsmath}
\usepackage{graphicx}
\usepackage{epsfig}
\usepackage{subfigure}
\usepackage{xcolor}

\begin{document}

\title{Quantum beats and metrology in a rapidly rotating Nitrogen-Vacancy center}

\author{Weijun Cheng}
\affiliation{Center for Quantum Sciences and School of Physics, Northeast Normal University, Changchun 130024, China}
\author{Tian Tian}
\affiliation{ School of Science, Changchun University, Changchun 130022, China}
\author{Zhihai Wang}
\email{wangzh761@nenu.edu.cn}
\affiliation{Center for Quantum Sciences and School of Physics, Northeast Normal University, Changchun 130024, China}

\begin{abstract}
In this paper, we study the dynamical behavior and quantum metrology in a rotating Nitrogen-Vacancy(NV) center system which is subject to an external magnetic field. Based on the recently realized rapid rotation of nano-rotor [J. Ahn, \textit{et. al.}, Phys. Rev. Lett. {\bf 121}, 033603 (2018) and  R. Reimann, \textit{et. al.}, Phys. Rev. Lett. {\bf 121}, 033602 (2018)], the frequency of the rotation is close to that of the intrinsic frequency of the NV center system, we predict the quantum beats phenomenon in the time domain and show that the quantum metrology can be enhanced by the superposition effect in our system.
\end{abstract}

\maketitle

\section{Introduction}
\label{Introduction}

The study of {Nitrogen-Vacancy (NV)} defect centers has increased steadily ever since the scanning confocal optical microscopy and magnetic resonance on  individual NV defect centers was achieved in 1997~\cite{11,12}.  Compared with the other spin system, NV center brings  many advantages: (1) Long coherence time even in the room temperature~\cite{31,32,33} which allows the system to accumulate signals over a long period of time and it is helpful to improve measurement accuracy~\cite{34}; (2) It is easy to initialize, operate and read~\cite{41}, and so on.

Therefore, the NV center in diamond~\cite{MW,RS} has attracted a lot of attentions both theoretically and experimentally. For instance, the temperature has an effect on  the longitudinal zero field splitting in NV center, which has been applied to  sensing biological cells~\cite{LP}, detection of temperature~\cite{VM,51,52,53,54}; the lateral stress and the electric field  lead to lateral  zero field splitting, so electric field~\cite{FD,71,72} or crystal strain~\cite{MW1} can be measured based on those features.  Additionally, the external magnetic field will cause the Zeeman splitting, so detecting the magnetic field~\cite{LR,61,62,63}  naturally adds to the potential applications of the NV center. An NV center in a mechanically rotating diamond has brought the sensing modalities into the rotating frame~\cite{AA} and it is found that the accumulated Berry phase can serve for the gyroscope~\cite{DM,MA,ZQY}. Also, an ensemble of NVs in rotating diamond have been used to explore the magnetic pseudofields generated in the rotating frame~\cite{AA1}.

We here study the dynamics of a NV center which is subject to a rotating magnetic field vector and a mechanical rotation. Our results show that the quantum beats phenomenon emerges with moderate magnetic field strength and a proper rotation velocity. {Furthermore, we also discuss the  behavior of the quantum Fisher information (QFI) with respective to the angular velocity. We find that the QFI will reach a relative large value in the parameter regime where the dynamics is characterized by different frequency components. In other words, the superposition effect is beneficial for enhancing the accuracy of sensing to the rotation velocity.}

The rest of the paper is organized as follows. In Sec.~\ref{Model}, we introduce the model and formulate the time evolution operator for the rotating NV center system. In Sec.~\ref{Beats}, we study the dynamics of the system and show that the population will exhibit a quantum beats phenomenon. In Sec.~\ref{Quantum Fisher information}, we study the behavior of QFI to the angular velocity. At last, we draw the conclusion in Sec.~\ref{results}.

\section{Model and Hamiltonian}
\label{Model}

Here, we apply the NV center to perform the quantum metrology on the gyroscope, that is, to estimate the angular velocity of the rotation.
Neglecting the strain-induced splitting, the Hamiltonian of {a} single NV center is written as ({hereafter}, we set $\hbar=1$ )
\begin{equation}
H=DS_{z}^{2}+g_e\mu_{B}\vec{B}\cdot\vec{S},
\label{1}
\end{equation}
where $D=(2\pi)2.87$ GHz is the zero field splitting between the $|m=0\rangle$ state and $|m=\pm1\rangle$ states, which will be shortened as $|0\rangle$ and $|\pm1\rangle$ in the rest of this paper. $S_{x},S_{y},S_{z}$ are the conventional Pauli spin-1 operators. $g_{e}=2$ is the ground-state Lande factor, $\mu_{B}=(2\pi)14$ MHz/mT is the Bohr magneton, and $\vec{B}$ is the magnetic field vector.

While a single NV center is placed in a rotating system with the rotating angular velocity $\Omega$, the Hamiltonian of NV center  becomes:
\begin{equation}
H(t)=DS_{z}'^{2}(t)+g_e\mu_{B}\vec{B}\cdot\vec{S'}(t).
\label{2}
\end{equation}
where $S_{z}'$ is associated with the rotation matrix $D_{\vec{n}}(\Omega t)$, with
\begin{eqnarray}
\left(
  \begin{array}{c}
  S_{x}'(t) \\
  S_{y}'(t) \\
  S_{z}'(t) \\
  \end{array}
\right)
&=&
D_{\vec{n}}(\Omega t)
\left(
  \begin{array}{c}
  S_{x} \\
  S_{y} \\
  S_{z} \\
  \end{array}
\right).
\label{3}
\end{eqnarray}
Here, $\vec{n}$ is  the direction vector of the rotation axis. For the sake of simplicity, we assume that the system is rotating counterclockwise about the $z$-axis. Then the rotation matrix $D_{\vec{n}}(\Omega t)$ can be  simplified to:
\begin{eqnarray}
D_{z}(\Omega t)&=&
\left(
  \begin{array}{ccc}
    \cos(\Omega t) & -\sin(\Omega t) & 0 \\
    \sin(\Omega t) & \cos(\Omega t) & 0 \\
    0 & 0 & 1 \\
  \end{array}
\right).
\label{4}
\end{eqnarray}
It's worth mentioning that {another} operation which is equivalent to $D_{\vec{n}}(\Omega t)$ satisfies the following relation:
\begin{eqnarray}
\vec{S'}(t)&=&R_{\vec{n}}^{\dag}(\Omega t)\vec{S}R_{\vec{n}}(\Omega t)
\label{5}
\end{eqnarray}
where $R_{\vec{n}}(\Omega t)=\exp(-i\Omega t\vec{n}\cdot\vec{S})$.
On the other hand, we consider that the system is subject to a magnetic field characterized by constant field strength and rotating along the $z$-axis, which can be written as:
\begin{eqnarray}
\vec{B}&=&B(\cos(\omega t)\vec{i}+\sin(\omega t)\vec{j})
\label{6}
\end{eqnarray}
with $B$ being the constant magnetic field strength.

It follows from Eqs. \eqref{2}-\eqref{4} and \eqref{6}  that the  time-dependent Hamiltonian for the rotating NV center in the rest reference frame is obtained as:
\begin{eqnarray}
H(t)&=&DS_{z}^{2}+g_e\mu_{B}B\{\cos(\Delta t)S_{x}-\sin(\Delta t)S_{y}\}.\nonumber\\
\label{7}
\end{eqnarray}
where $\Delta=\Omega-\omega$.
Therefore,  it is equivalent to a magnetic field rotating around the $z$-axis counterclockwisely with the effective angular velocity of $\Delta$. In such a scheme, we can measure $\Delta$ to estimate angular velocity $\Omega$ of the NV center. Moreover, the parameter $\Delta$ is introduced into the dynamics of the system due to the constant magnetic field strength $B$.

Assuming the NV center at an arbitrary moment is described by the wave function:
 {\begin{eqnarray}
|\psi(t)\rangle=a(t)|1\rangle+b(t)|0\rangle+c(t)|-1\rangle,
\label{state}
\end{eqnarray}}
where $a(t)$, $b(t)$ and $c(t)$ are the amplitudes for finding the system in the state $|1\rangle$, $|0\rangle$ and $|-1\rangle$, respectively.
 The Schrodinger's equation in the rest reference frame is written as:
 \begin{eqnarray}
i\frac{\partial}{\partial t}
\left(
  \begin{array}{c}
    a(t) \\
    b(t) \\
    c(t) \\
  \end{array}
\right)
&=&
\left(
  \begin{array}{ccc}
    D & Ae^{i\Delta t} & 0 \\
    Ae^{-i\Delta t} & 0 & Ae^{i\Delta t} \\
    0 & Ae^{-i\Delta t} & D \\
  \end{array}
\right)
\left(
  \begin{array}{c}
    a(t) \\
    b(t) \\
    c(t) \\
  \end{array}
\right),\nonumber\\
\label{8}
\end{eqnarray}
 where $A=g_e\mu_{B}B/\sqrt{2}$.
 With the help of the transformation:
 \begin{eqnarray}
a(t)&=&\bar{a}(t)\exp(-iDt)\exp(i\Delta t),\nonumber\\
b(t)&=&\bar{b}(t)\exp(-iDt),\nonumber\\
c(t)&=&\bar{c}(t)\exp(-iDt)\exp(-i\Delta t),
\label{9}
\end{eqnarray}
the Schrodinger's equation will turn into another form:
 \begin{eqnarray}
i\frac{\partial}{\partial t}Y&=&MY,
\label{10}
\end{eqnarray}
 where
  \begin{eqnarray}
Y&=&(\bar{a}(t), \bar{b}(t), \bar{c}(t))^{\top}
\label{state1}
\end{eqnarray}
and
  \begin{eqnarray}
M&=&\left(
  \begin{array}{ccc}
    \Delta & A & 0 \\
    A & -D & A \\
    0 & A & -\Delta \\
  \end{array}
\right),
\label{state1}
\end{eqnarray}
is a time-independent matrix.

\section{Fourier spectral-decomposition and quantum beats}
\label{Beats}

In the above section, we have outlined the dynamical process for a single NV center yielding an external magnetic field and a mechanical rotation.
 It is reported recently that the rotation frequency of a {nano-rotor} can be achieved by 1GHz~\cite{JA,RR}, which is in the same order of the energy-level spacing of our system. Naturally, during the evolution of our linear system, multiple frequency components will be represented, which will induce a quantum beat phenomenon under the appropriate parameters. In this section, {we will present the time evolution of our system within diverse parameters.}

 Considering the nowadays experimental feasibility, we prepare the system in the state $|0\rangle$ via the optical pumping technology initially.
 {The dynamical evolution can be calculated by Eq.~\eqref{10}, but the analytical results are cumbersome and we will present the numerical results in what follows.} The population of the system in the state $|0\rangle$ is then obtained as:
\begin{eqnarray}
P(t)&=&|b(t)|^{2}=|\bar{b}(t)|^{2}.
\label{12}
\end{eqnarray}
 To investigate the dynamics in a detailed way, we here perform the numerical  Fourier transformation from the time domain to the frequency domain.
  It then yields
  \begin{eqnarray}
\bar{P}(\omega)&=&\frac{1}{\sqrt{2\pi}}\int_{-\infty}^{\infty}P(t)e^{-i\omega t}dt\nonumber \\
&=&\frac{1}{\sqrt{2\pi}}\left[\begin{array}{c}
K_{0}\delta(\omega)
+\sum_{i=1}^{3}K_{i}\delta(\omega-\omega_{i})
\end{array}\right].\nonumber\\
\label{13}
\end{eqnarray}

 \begin{figure}[tbp]
\centering
\includegraphics[width=8cm]{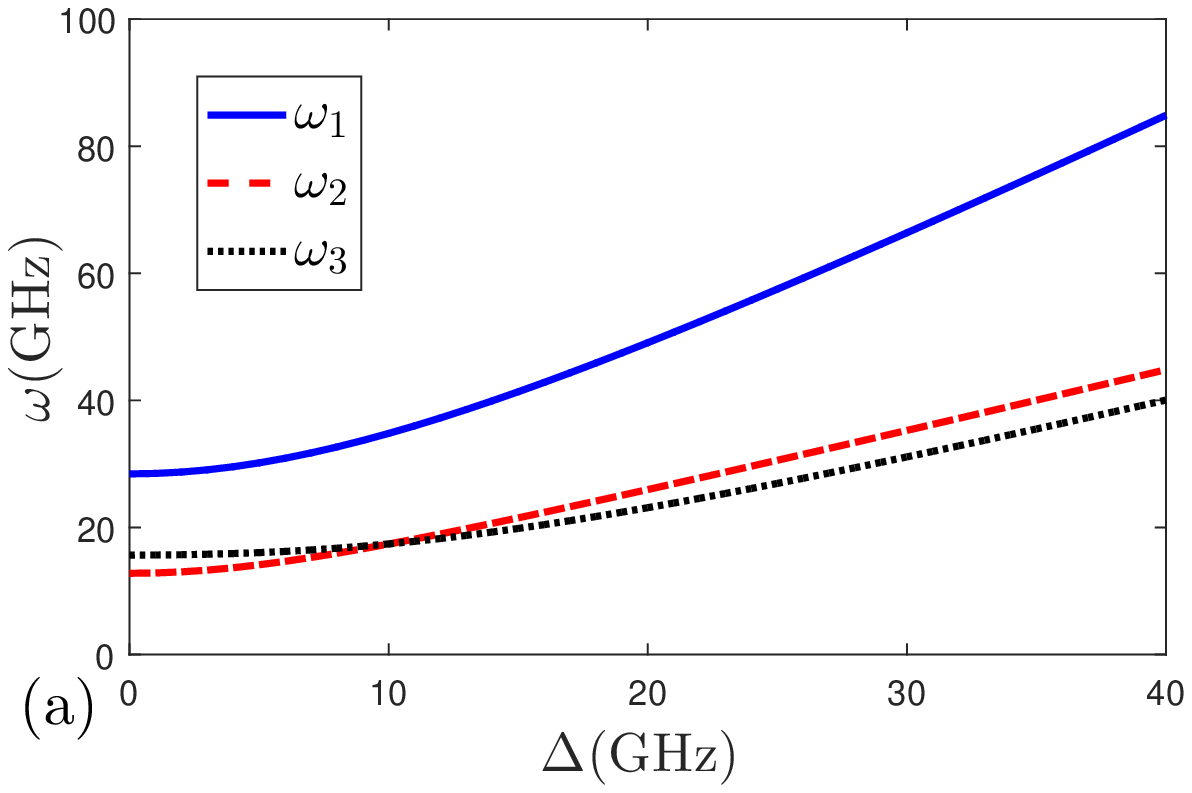}
\includegraphics[width=8cm]{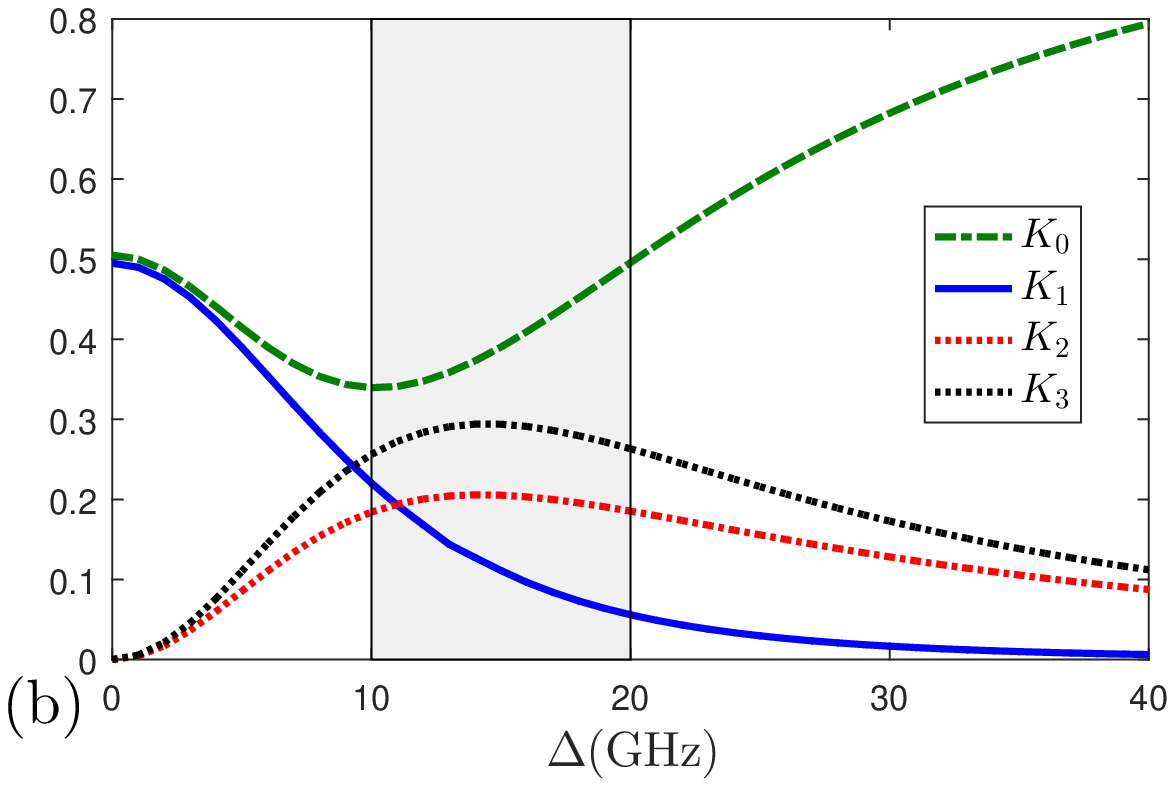}
\caption{(Color online) (a) The frequency components $\omega_i\,(i=1,2,3)$ versus $\Delta$. (b) The amplitudes $K_{0}$, $K_{1}$,  $K_{2}$ and $K_{3}$ for different frequency components versus $\Delta$. The parameters are set as $A=10$GHz.}
\label{NVF}
\end{figure}

\begin{figure}[tbp]
\begin{center}
\includegraphics[width=8cm]{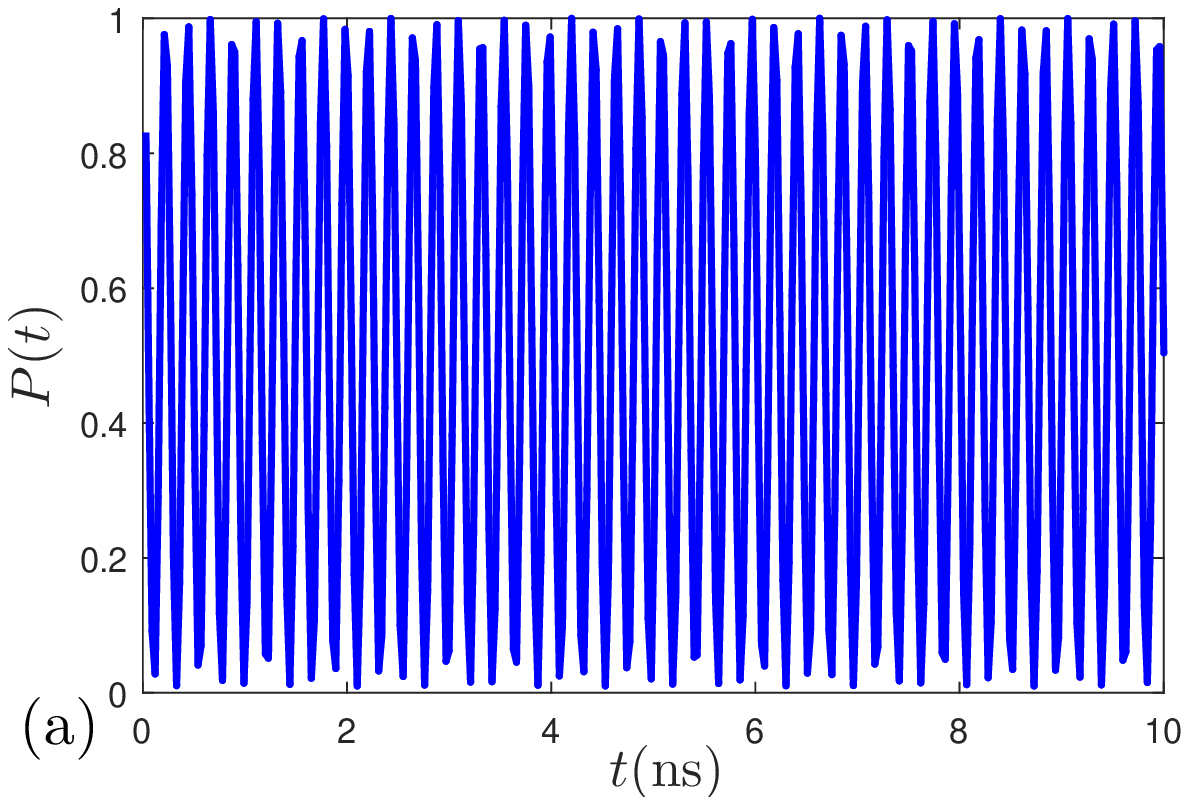}
\includegraphics[width=8cm]{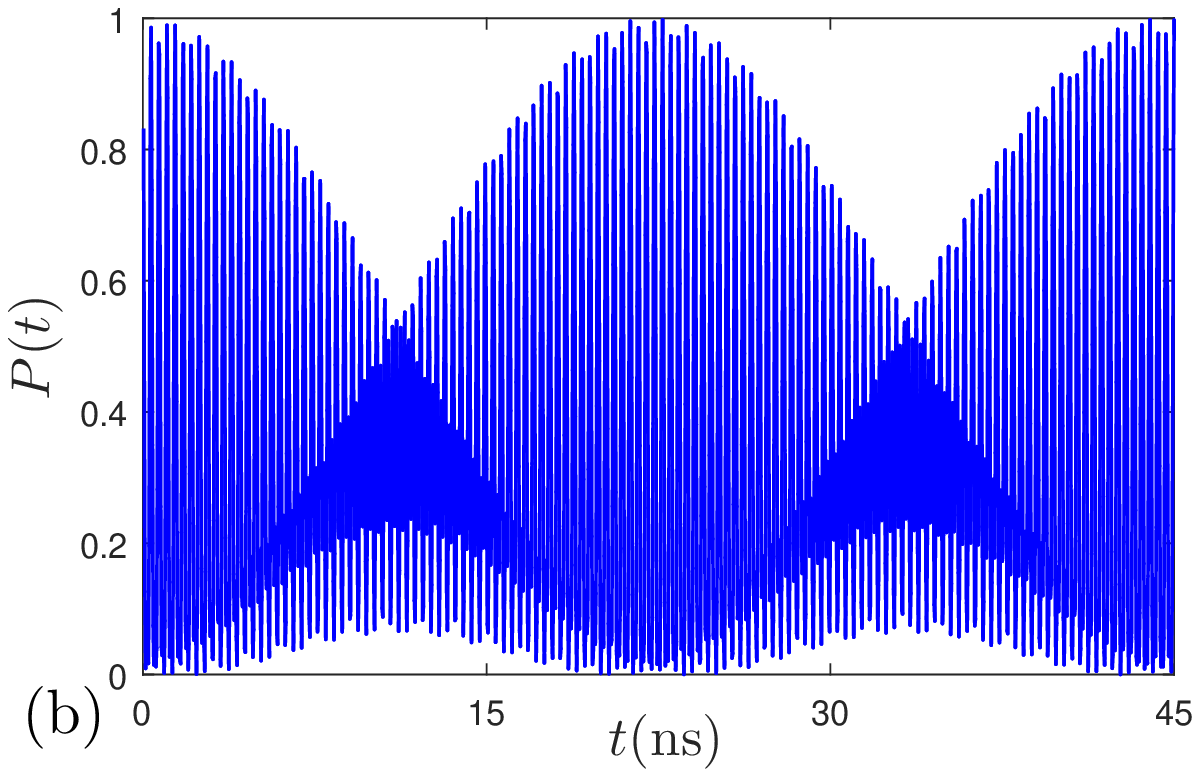}
\includegraphics[width=8cm]{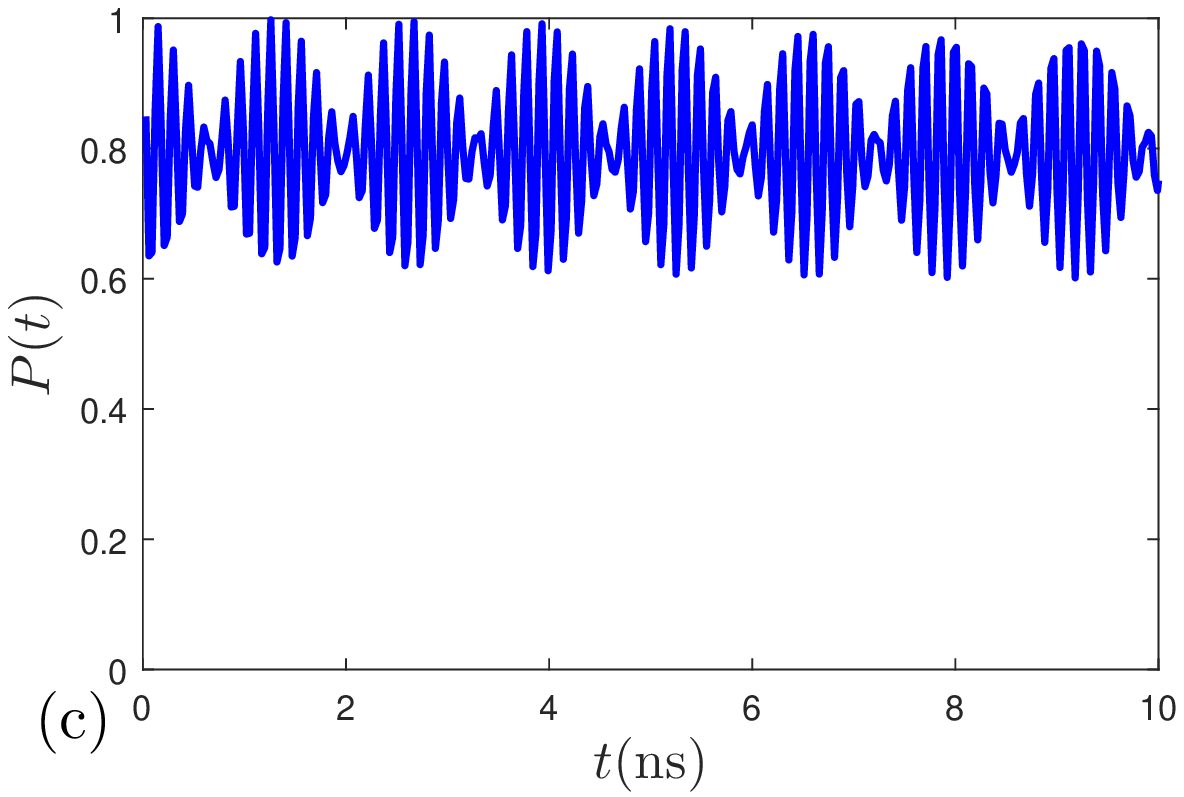}
\end{center}\vspace{-0.5cm}
\caption{(Color online) The probability $P(t)$ versus the time $t$. The parameters are set as $A=10$GHz. (a)$\Delta=0$, (b)$\Delta=10.8$ GHz, (c)$\Delta=40$ GHz.}
\label{D}
\end{figure}

In Fig.~\ref{NVF}(a), we plot the curves of the different frequency components $\omega_{i}\,(i=1,2,3)$  versus $\Delta$. Here, for the sake of simplicity, we set the strength of magnetic field as $B=5\sqrt{2}/14$ T, so as $A=10$ GHz. Moreover, in Fig.~\ref{NVF}(b), we also plot the curves of amplitudes $K_{i}(i=0-3)$ for different frequency components versus $\Delta$. It is clear that the evolution of the system will only show one frequency $\omega_1$ at $\Delta=0$, that is, the effective magnetic field is at the $x$-axis. For a finite $\Delta$, there will be three frequency components in our system. In particular, when $\Delta \approx 10.8$ GHz, two of the above three frequencies will be overlapped  in such situations. In addition, when $\Delta$ is big enough,  the course of evolution will cover two frequencies that have a similar amplitude to each other.

To show the combining effect of the oscillations with different frequencies, we plot the dynamical behavior $P(t)$ as a function of the evolution time in Fig.~\ref{D}. As shown in Fig.~\ref{D}(a), when $\Delta=0$, it shows a Rabi oscillation with a single frequency, which is coincidence with Fig.~\ref{NVF}(b) in that $K_{0}=K_{1}$ and $K_{2}=K_{3}=0$. For a moderate $\Delta$($\Delta\approx10.8$ GHz), there is two frequency components with equal frequency and large amplitude difference, the dynamics is illustrated in Fig.~\ref{D}(b). As the further increase of  $\Delta$, the two participated frequency components approach each other and it behaves as a beat phenomenon as show in Fig.~\ref{D}(c). The quantum beats have been also predicted and observed in an ensemble of $\Lambda$-type three-level atoms with nearly degenerate excited-states and other quantum systems~\cite{MOS,WW,RH,TH,MM1,GC,TL,DG}. Moreover, the quantum beats have been reported to be observed in NV center system, which is subject to a periodical modulation~\cite{XFHe,SC,KF}.

{The optically detected magnetic resonance approach supplies us a method to track the population evolution of NV center triplet ground state. In the experiments, people usually apply the laser to pump the NV center from the ground state to its excited states, which are also triplet and collect the fluorescence emitted by spontaneous emissions~\cite{MW}. A recent research shows that the $100$ ps laser pulse for the pumping can be realized~\cite{laser}, and therefore, the quantum beats phenomenon is expected to be observed.}

\section{Quantum Fisher information}
\label{Quantum Fisher information}

In the above section, we have illustrated  the evolution of our system and analyzed their frequency spectrum.  Motivated by the potential application for sensitive gyroscope based on our system, we will discuss the accuracy of the parameter estimation by studying the QFI with respective to the angular velocity $\Omega$.

The QFI is a central quantity in the field of quantum metrology and quantum estimation theory. It is introduced by extending the classical Fisher information to the quantum regime, and characterizes how sensitive of a parameter estimation can be achieved by use of the quantum source in a system. According to the quantum Cramer-Rao inequality, the uncertainty $\delta x$ in the estimation for the physical parameter $x$  is bounded by $\delta x\geq1/\sqrt{\nu\mathcal{F}_{x}}$, where $\nu$ is the times of the independent measurements and $\mathcal{F}_{x}$ is the QFI. A larger QFI corresponds to a more accurate estimation to the parameter~\cite{JA,RR,SL1,SL2}.

For a general quantum state which is described by the density matrix $\rho$, with the spectral decomposition $\rho=\sum_{i=1}^{M}p_{i}|\psi_{i}\rangle\langle\psi_{i}|$
(where $M$ denotes the number of nonzero $p_{i}$), the QFI is given by~\cite{YM,WZ,JL,JL1}
\begin{eqnarray}
\mathcal{F}_{x}&=&\sum_{i=1}^{M}\frac{(\partial_{x}p_{i})^{2}}
{p_{i}}+4\sum_{i=1}^{M}p_{i}\langle\partial_{x}\psi_{i}|\partial_{x}
\psi_{i}\rangle\nonumber \\&&-\sum_{i,j=1}^{M}\frac{8p_{i}p_{j}}{p_{i}+p_{j}}|\langle\psi_{j}|
\partial_{x}\psi_{i}\rangle|^{2}.
\label{14}
\end{eqnarray}
 In this paper, we input a pure state $|0\rangle$ initially, according to the unitary evolution governed by Eq.~\eqref{10}, the final state is also a pure state. As a result, the QFI in Eq.~\eqref{14} can be further reduced to
 \begin{equation}
 \mathcal{F}_{x}=4(\langle\partial_{x}\psi|\partial_{x}
 \psi\rangle-|\langle\psi|\partial_{x}\psi\rangle|^{2}).
 \label{15}
 \end{equation}

 \begin{figure}[tbp]
\centering
\includegraphics[width=8cm]{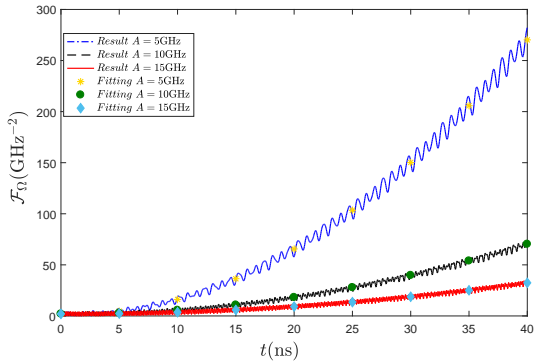}
\caption{(Color online) The QFI $\mathcal{F}_{\Omega}$ versus the time $t$. The parameters are set as $\Delta=20$ GHz.}
\label{QFIt}
\end{figure}
\begin{figure}[tbp]
\centering
\includegraphics[width=8cm]{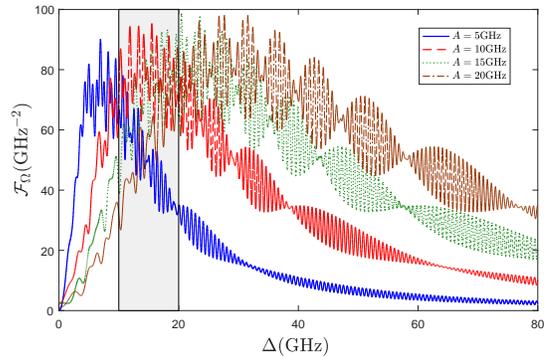}
\caption{(Color online) The QFI $\mathcal{F}_{\Omega}$ versus $\Delta$, The parameters are set as $t=10$ ns.}
\label{QFIthta}
\end{figure}
Now, we investigate the QFI to the angular velocity $\Omega$, motivated by the potential application in quantum gyroscope. Choosing the parameter $A=5\,(10, 15)$ GHz, we numerically obtain the QFI to $\Omega$, and demonstrate it in Fig.~\ref{QFIt}.  It obviously behaves as a concussive growth function with the evolution time. Furthermore, numerical fitting results are also exhibited by the dotted lines.
The data fittings show that they yield  quadratic forms as:
\begin{eqnarray}
\mathcal{F}_{\Omega}(t)& \approx &0.1764t^{2}-0.3514t+2.0064, A=5\,\rm{GHz};\nonumber \\
\mathcal{F}_{\Omega}(t)& \approx &0.0449t^{2}-0.0888t+2.0013,  A=10\,\rm{GHz};\nonumber \\
\mathcal{F}_{\Omega}(t)& \approx &0.02t^{2}-0.0404t+2.0082, A=15\,\rm{GHz}.
 \label{16}
\end{eqnarray}
In other words, the QFI $\mathcal{F}_{\Omega}(t)$ shows an approximation of quadratic functions.

Furthermore,  for a fixed time $t=10$ ns, the QFI to $\Omega$ is plotted as a function of $\Delta$ in Fig.~\ref{QFIthta}, where the blue (red, green, brown) line is plotted at $A=5\,(10,15,20)$ GHz. In the gray region of the figure, we observe that the QFI are relatively larger than other regions. {We also observe from  Fig.~\ref{D}(b) that the non-zero frequency component owns relatively large amplitudes in  this regime. Therefore, the superposition of non-zero frequency oscillation is beneficial for parameter estimation.}

\section{Results and discussions}

\label{results}

In this paper, we have investigated the dynamics and quantum metrology in a NV center system which is subject to a rotation and external magnetic field vector.  Benefited from the realization of rapid rotation of nano-mechanical rotor, the frequency of the rotation is achieved by GHz, which is in the same order with the intrinsic characteristic frequency of the NV center. We find that the superposition of oscillations will enhance the accuracy of parameter estimation.

{In our scheme, the magnetic field is rotated around the same axis as the mechanical rotation of the NV center. As a result, the system is equivalent to be subject to an effective magnetic field with modified frequency. Our results show that, the QFI can be enhanced by adjusting the amplitudes of the magnetic field. We hope that our metrology scheme with the assistance of tunable magnetic field can be useful for the designing of gyroscope based on solid state spin system.}

\noindent \textbf{Acknowledgments.}
 This work is supported by the National Natural Science Foundation of China  (Grant No. 11875011), Jilin province science and technology development plan item (Grant No. 20180520175JH),  Educational Commission of Jilin Province of China (Grant No. JJKH20190266KJ) , and the Project of cultivating young teachers in Changchun University (Grant No. ZK201809).


\begin{thebibliography}{99}

\bibitem{11} A. Gruber, A. Drbenstedt, C. Tietz, L. Fleury, J. Wrachtrup, C. Von Borczyskowski, Science {\bf 276}, 2012  (1997).

\bibitem{12} D.-D. Awschalom, M. E. Flatt¨¦, Nat. Phys. {\bf 3}, 153 (2007).

\bibitem{31} N. Zhao, S.-W. Ho,  R.-B. Liu, Phys. Rev. B {\bf 85}, 115303 (2012).

\bibitem{32} L. Viola, E. Knill,  S. Lloyd, Phys. Rev. Lett. {\bf82}, 2417 (1999).

\bibitem{33}  L. Cywi\'{n}ski, R.-M. Lutchyn, C.-P. Nave, S.-D. Sarma, Phys. Rev. B {\bf77}, 174509 (2008).

\bibitem{34} Y.-N. Fang, X. Xiao, C.-P. Sun, W. Yang, N. Zhao,  arXiv:1807.02644 (2018).

\bibitem{41} J.-M. Cai, B. Naydenov, R. Pfeiffer, L.-P. McGuinness, K.-D. Jahnke, F. Jelezko, M.-B. Plenio,  A. Retzker, New J. phys. {\bf14}, 113023 (2012).

\bibitem{MW} M.-W. Doherty, N.-B. Manson, P.-Delaney, F.-Jelezko, J.-Wrachtrup,  L.-C.-L. Hollenberg, Phys. Rep. {\bf 528}, 1 (2013).

\bibitem{RS}  R. Schirhagl, K. Chang, M. Loretz,  C.-L. Degen,  Ann. Rev. Phys. Chem. {\bf 65}, 83 (2014).

\bibitem{LP} L. P. McGuinness, Y. Yan, A. Stacey, D. A. Simpson, L. T. Hall, D. Maclaurin, S. Prawer, P. Mulvaney, J. Wrachtrup, F. Caruso, R. E. Scholten,  L. C. L. Hollenberg,  Nat. Nanotechnol. {\bf 6}, 358 (2011).

\bibitem{VM} V. M. Acosta, E. Bauch, M. P. Ledbetter, A. Waxman, L.-S. Bouchard,  D. Budker, Phys. Rev. Lett. {\bf 104}, 070801 (2010).

\bibitem{51}  G. Kucsko, P.-C. Maurer,  N.-Y. Yao, M. Kubo,  H.-J. Noh,  P.-K. Lo, H. Park,  M.-D. Lukin, Nature {\bf500}, 54 (2013).

\bibitem{52}  D.-R. Glenn,  K. Lee,  H. Park,  R. Weissleder,  A. Yacoby,  M.-D. Lukin, H. Lee,  R.-L. Walsworth,  C.-.B Connolly, Nat. Methods {\bf12}, 736 (2015).

\bibitem{53} D.-L. Sage, K. Arai, D.-R. Glenn, S.-J. DeVience, L.-M. Pham, L.-R. Lee, M.-D. Lukin, A. Yacoby, A. Komeili, R.-L. Walsworth,  Nature {\bf496}, 486 (2013).

\bibitem{54} F. Shi, Q. Zhang, P. Wang, H. Sun, J. Wang, X. Rong, M. Chen, C. Ju, F. Reinhard, H. Chen, J. Wrachtrup, J. Wang,  J. Du, Science {\bf347}, 1135 (2015).

\bibitem{FD} F. Dolde, H. Fedder, M.-W. Doherty, T. Nobauer, F. Rempp, G. Balasubramanian, T. Wolf, F. Reinhard, L.-C. L. Hollenberg, F. Jelezko,  J. Wrachtrup,  Nat. Phys. {\bf7}, 459(2011).

\bibitem{71} Y. Martin, D.-W. Abraham, H.-K. Wickramasinghe, Appl. Phys. Lett. {\bf52}, 1103 (1988).

\bibitem{72} F. Jelezko, J. Wrachtrup, Phys. Status Solidi A {\bf203}, 3207 (2006).

\bibitem{MW1} M. W. Doherty, V. V. Struzhkin, D. A. Simpson, L.-P.-M. Guinness, Y. Meng, A. Stacey, T. J. Karle, R. J. Hemley, N. B. Manson, L.-C.-L. Hollenberg, S. Prawer, Phys. Rev. Lett. {\bf 112}, 047601 (2014).

\bibitem{LR} L. Rondin, J.-P. Tetienne, T. Hingant, J.-F. Roch, P. Maletinsky,  V. Jacques, Rep. Prog. Phys. {\bf 77}, 056503 (2014).

\bibitem{61} G. Balasubramanian, P. Neumann, D. Twitchen, M. Markham, R. Kolesov, N. Mizuochi, J. Isoya, J. Achard, J. Beck, J. Tissler, V. Jacques, P.-R. Hemmer, F. Jelezko, J. Wrachtrup, Nat. Mater. {\bf 8}, 383(2009).

\bibitem{62} T. Wolf, P. Neumann, K. Nakamura, H. Sumiya, T. Ohshima, J. Isoya, J. Wrachtrup, Phys. Rev. X {\bf 5}, 041001 (2015).

\bibitem{63} G. Balasubramanian, I.-Y. Chan, R. Kolesov, M.-A. Hmoud, J. Tisler, C. Shin, C. Kim, A. Wojcik, P.-R. Hemmer, A. Krueger, T. Hanke, A. Leitenstorfer, R. Bratschitsch, F. Jelezko, J. Wrachtrup, Nature {\bf 455}, 648 (2008).

\bibitem{AA} A. A. Wood, E. Lilette, Y. Y. Fein, N. Tomek, L. P. McGuinness, L. C. L. Hollenberg, R. E. Scholten, A. M. Martin, Sci. Adv. {\bf 4}, eaar7691 (2018).

\bibitem{DM} D. Maclaurin, M. W. Doherty, L. C. L. Hollenberg, A. M. Martin,  Phys. Rev. Lett. {\bf 108}, 240403 (2012).

\bibitem{MA} M. A. Kowarsky, L. C. L. Hollenberg, A. M. Martin,  Phys. Rev. A {\bf 90}, 042116 (2014).

\bibitem{ZQY} {X. Y. Chen, T. Li, Z. Q. Yin, Science Bulletin {\bf64}, 380 (2019).}

\bibitem{AA1}A. A. Wood, E. Lilette, Y. Y. Fein, V. S. Perunicic, L. C. L. Hollenberg, R. E. Scholten, A. M. Martin,  Nat. Phys. {\bf 13}, 1070 (2017).

\bibitem{MOS} M. O. Scully,  M. S. Zubairy, \textit{Quantum Optics} (Cambridge
university press, 1997).

\bibitem{WW}W. Chow, M. O. Scully, J. Stoner, Phys. Rev. A {\bf 11}, 1380 (1975).

 \bibitem{RH}R. Herman, H. Grotch, R. Kornblith, J. Eberly, Phys. Rev. A {\bf 11}, 1389 (1975).

 \bibitem{TH}T. H. Jeys, F. B. Dunning, R. F. Stebbings, Phys. Rev. A {\bf 29}, 379 (1984).

\bibitem{MM1} M. Mitsunaga, C. L. Tang, Phys. Rev. A {\bf 35}, 1720 (1987).

\bibitem{GC}G. C. Hegerfeldt, M. B. Plenio, Phys. Rev. A {\bf 47}, 2186 (1993).

\bibitem{TL}T. Legero, T. Wilk, M. Hennrich, G. Rempe,  A. Kuhn, Phys. Rev. Lett. {\bf 93}, 070503 (2004).

\bibitem{DG}D. G. Norris, L. A. Orozco, P. B.-Blostein,  H. J. Carmichael, Phys. Rev. Lett. {\bf 105}, 123602 (2010).

\bibitem{XFHe}X.-F. He , P. T. H. Fisk , N. B. Manson, J. Lumin. \textbf{60$\&$61}, 739 (1994).

\bibitem{SC}S. C. Rand, A. Lenef, S. W. Brown, J. Lumin. {\bf 53}, 68 (1992).

\bibitem{KF}K. Fang, V. M. Acosta, C. Santori,  Z. Huang,  K. M. Itoh,  H. Watanabe,
S. Shikata, R. G. Beausoleil,  Phys. Rev. Lett. {\bf 110}, 130802 (2013).

\bibitem{laser}{L. Hacquebard, L. Childress, Phys. Rev. A {\bf 97}, 063408 (2018).}

\bibitem{JA} J. Ahn, Z. Xu,  J. Bang, Y.-H. Deng,  T. M. Hoang, Q.  Han,
R.-M. Ma,  T. Li,  Phys. Rev. Lett. {\bf 121}, 033603 (2018).

\bibitem{RR} R. Reimann, M. Doderer, E. Hebestreit, R. Diehl, M.
Frimmer, D. Windey, F. Tebbenjohanns, L. Novotny,
Phys. Rev. Lett. {\bf 121}, 033602 (2018).

\bibitem{SL1} S. L. Braunstein, C. M. Caves, Phys. Rev. Lett. {\bf 72}, 3439 (1994).

\bibitem{SL2} S. L. Braunstein, C. M. Caves, G. J.  Milburn,  Ann. Phys. NY {\bf 247}, 135 (1996).

\bibitem{YM}Y. M. Zhang, X. W. Li, W. Yang, G. R. Jin,  Phys. Rev. A {\bf 88}, 043832 (2013).

\bibitem{WZ}W. Zhong, Z. Sun, J. Ma, X. Wang, F. Nori, Phys. Rev. A {\bf 87}, 022337 (2013).

\bibitem{JL} J. Liu, H.-N. Xiong,  F. Song, X. Wang,  Physica A {\bf 410}, 167(2014).

\bibitem{JL1}  J. Liu,  X.-X. Jing, W. Zhong, X.-G. Wang,  Commun. Theor. Phys. {\bf 61}, 45 (2014).



\end{thebibliography}
\end{document}